\documentstyle[prl,aps,multicol,epsfig]{revtex}
\begin{document}
\draft
\widetext
\title{  Pairing  due to Spin Fluctuations in 
Layered Organic Superconductors }
\author{J\"org Schmalian    }
\address{ University of Illinois at 
Urbana-Champaign,   Loomis Laboratory 
of Physics, 1110 W. Green,  Urbana, IL, 61801 }
\date{ \today}
\maketitle
\widetext 
\begin{abstract} 
\leftskip 54.8pt
\rightskip 54.8pt
 I  show that for a $\kappa$-type organic (BEDT-TTF)$_2$X molecular
crystal, a superconducting state with $T_{\rm c} \approx
10\, {\rm K}$ and 
  gap   nodes  on the Fermi surface can be 
     caused    by    short-ranged antiferromagnetic   spin fluctuations.   
Using a  two-band description for the anti-bonding  orbitals
on a  BEDT-TTF dimer of  the  $\kappa$-type  salt, 
and an intermediate local Coulomb  repulsion between two holes on  one
dimer,  the magnetic interaction  and
the superconducting gap-function are determined self consistently
within  the fluctuation exchange approximation. 
The pairing interaction is predominantly caused by inter-band coupling
and additionally affected by  spin excitations of   the 
quasi one-dimensional band.
\end{abstract}   
\begin{multicols}{2}
\narrowtext   
Electronic mechanisms for   superconductivity are of conceptual
interest on  their  own and    allow, in principle,     large
superconducting (SC) transition temperatures since  electronic
energy scales   easily exceed phononic energies.
In addition to the  copper-oxide based high-temperature superconductors, 
heavy fermion superconductors like UPt$_3$ and UBe$_{13}$, and 
 the ruthenate compound Sr$_2$RuO$_4$,
organic superconductors are   candidates for   a  
pairing mechanism which originates in strong electronic correlations
in the normal state.

The  organic molecular crystals $\kappa$-(BEDT-TTF)$_2$X
 (in the following abbreviated as  $\kappa$-(ET)$_2$X), 
are   very anisotropic quasi 
two-dimensional superconductors with  transition temperatures 
up to around $10\,
  {\rm K}$, depending on pressure and the ion X, which
  can, for example, be I$_3$,
Cu[N(CN)$_2$]Br   or Cu(SCN)$_2$~\cite{Lang}.
Shubnikov - de Haas experiments have established the existence of a well
defined Fermi surface in these materials~\cite{Caul94,Mie97},
 demonstrating  the Fermi liquid character of the 
low energy quasiparticles.
Below T$_c$,  the low temperature  $^{13}$C-NMR spin-lattice
 relaxation rate   varies as $T^3$~\cite{DSS95,May95,Kan96}, 
 the electronic specific heat coefficient exhibits a
magnetic  field dependence  
 $\gamma \propto B^{1/2}$~\cite{NK97} and the thermal conductivity $\kappa$
 is linear in $T$~\cite{BBD98}; evidences  for
   nodes of the SC gap function on the Fermi surface.
All these results suggest   that   superconductivity is   caused by a highly
anisotropic and likely  non-phononic pairing interaction.
An electronic mechanism is also  supported by the absence
 of an isotope shift of $T_c$
due to the C=C and C-S phonon modes, as observed by $^{34}$S
and $^{13}$C=$^{13}$C isotope substitution in the central part of the 
ET-molecule~\cite{KB95}.

 It has      been pointed out earlier 
 that many  anomalous normal state properties of  layered organics 
 strongly suggest  the important role of    
  electronic correlations~\cite{Bul88,KF96,KK97}
and that some, but not all,  of these anomalies are in striking
similarity to cuprate 
superconductors~\cite{RHM97}.
Particularly  magneto-transport, the thermoelectric power, and the
uniform magnetic susceptibility  display similar behavior.
Another resemblance is   their proximity to an antiferromagnetic (AF) state,
even though,  in organics, frustration and strong
in-plane anisotropies  demonstrate    differences between both systems.
Particularly, due to the, by now   established,  dominant
d$_{x^2-y^2}$ order parameter  component~\cite{wollman}, high-temperature 
 superconductivity in   cuprates is believed to be 
dominated or entirely caused  by an electronic mechanism.
 The spin fluctuation scenario~\cite{DP,DJS}, which   predicted 
the  d$_{x^2-y^2}$ state, is a very promising approach,   because it also 
offers  an explanation for a variety of anomalous normal state properties
of cuprates.
 The unfrustrated  AF  nearest
neighbor  coupling  and the specific shape of the Fermi surface are
important  prerequisites for  a spin-fluctuation induced 
SC state with  $d_{x^2-y^2}$ symmetry
in cuprates.
In contrast, in organics, the magnetic coupling is frustrated, due to the
underlying anisotropic triangular lattice~\cite{RHM97}. 
In addition,    their Fermi surface consists
of two  disconnected  pieces~\cite{Caul94,Mie97}.
 Therefore, it is not at all  obvious whether, despite these numerous
 similarities  to  cuprates,  spin fluctuations 
  can  bring about     superconductivity in  the
 organics.

In this paper I  show that  spin fluctuations can cause 
superconductivity with    $T_c \approx 10-15 \, {\rm K}$
in $\kappa$-type organic molecular crystals~\cite{alpha}
and discuss similarities between the inter-band coupling and
the {\em hot spot} scenario of cuprates~\cite{DP,HR}.
Gap nodes on the Fermi surface,  accompanied 
 by  changes in sign of the order parameter,  result
because of    zone boundary 
inter-band pairing combined with  a
  intra-band coupling of a quasi one-dimensional
(quasi 1-D) band. This  is in
agreement with NMR and specific heat measurements
 and should be observable in a corresponding phase shift experiment.
Furthermore,  the strong inter-band pairing interaction
 causes  an  opposite sign of the gap functions of the two
bands.

\begin{figure}
\centerline{\epsfig{file=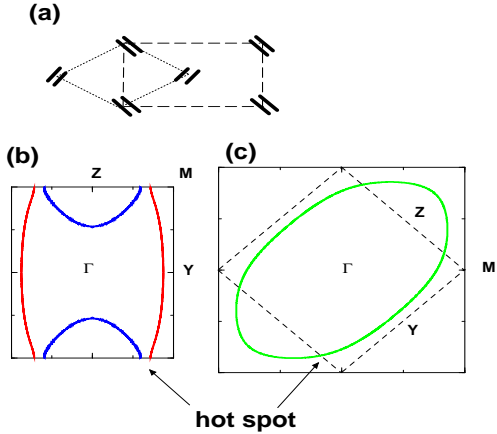,width=8.cm, height=6cm, scale=0.85}}
\caption{(a) Spatial arrangement of the ET molecules within the planes.
The  two dimer types are indicated by the different tilting of the molecule pairs,
which are, after dimerization,   considered as a single electronic degree of freedom.
The dashed (dotted) lines indicate the unit cell for $t_3 \neq t_{3'}$ ($t_3 = t_{3'}$).
(b) Fermi surface for the case $t_3 \neq t_{3'}$, consisting of  a 
quasi 1-D band 
and  a hole pocket closed around the $Z$-point.
(c) Fermi surface for the case $t_3 = t_{3'}$. Note that the {\em hot spots}
due to  nearest neighbor AF correlations correspond to
the gaps between both Fermi surface parts in (b). } 
\label{fig1}
\end{figure}

Within a given conducting layer, the
 unit cell of $\kappa$-(ET)$_2$X is occupied with six electrons in the
outer  electronic shells and consists of  four ET- molecules, which can be
considered as single electronic degrees of freedom~\cite{OMI88}.
The four molecules
are arranged in two dimers (see Fig.~\ref{fig1}a).
 Due to the large intra-dimer transfer
integral, $t_1 \approx 250\, {\rm meV}$, the
two bonding bands are completely occupied and shifted to binding energies
$\sim 2t_1$ below the Fermi energy.
Only the two remaining   anti-bonding bands cross the Fermi surface;
 those  are half filled  with the remaining two electrons.
\noindent
In  what  follows, I  consider  only the two anti-bonding 
bands   near the Fermi surface (within a hole picture) and neglect the bonding bands.
Due to the strong intra-molecular 
Coulomb repulsion ($U_{\rm ET} \geq 1 \, {\rm eV}$), a
 doubly occupied dimer will  
likely distribute the two holes on both molecules of the dimer, 
causing an effective correlation energy $U \sim 2 t_1$ 
determined by the bonding-anti-bonding band splitting~\cite{RHM97}.
This leads   to the following two band Hubbard Hamiltonian 
\begin{eqnarray}
H&=&\sum_{ij,l,\sigma} \, t_{ij} c^\dagger_{il\sigma} c_{jl\sigma}
+ U \sum _{i,l} n_{il\uparrow} n_{il\downarrow} \nonumber \\
& &+\sum_{ij,\sigma} \, \tilde{t}_{ij} \left(c^\dagger_{i1\sigma} c_{j2\sigma}
+c^\dagger_{j2\sigma} c_{i1\sigma} \right)\, .
\label{ham}
\end{eqnarray}
Here, $c^\dagger_{il\sigma} $ is the creation operator of a hole with spin 
$\sigma$ in the anti-bonding band of the $l$-th dimer ($l \in \{1,2\}$)
in the $i$-th unit cell and $n_{il\sigma} = 
c^\dagger_{il\sigma}c _{il\sigma}$.
$t_{ij}$ and $\tilde{t}_{ij}$ are   inter-dimer hopping elements
 between dimers of the same and of different type, respectively. Their 
Fourier transforms $t_{\bf k}$ and $\tilde{t}_{\bf k}$
 with ${\bf k}=(k_y,k_z)$   determine
the band-structure of the two bands
 $\varepsilon_{{\bf k}, \pm} = t_{\bf k} \pm \tilde{t}_{\bf k}$.
 Various different tight binding parameterizations have been proposed for
$t_{\bf k}$ and $\tilde{t}_{\bf k}$~\cite{Caul94,OMI88,Tam91,IYU97}.
I will use the dispersion relation $t_{\bf k}=2 t_2 \cos(k_y)$
and $\tilde{t}_{\bf k}= 2\cos(k_z/2)\sqrt{t_3^2+t_{3'}^2+2t_3 t_{3'}
\cos(k_y)}$~\cite{Caul94,IYU97}, which can be derived from a four 
 band  model  up to leading order in $t_2/t_1$.
Due to the non-symmetrical position of the dimers 
and the ET-molecule structure itself, the three hopping elements
  are slightly different.
The Fermi surface, for $t_2=45 \, {\rm meV}$, 
$t_3=60 \, {\rm meV}$ and $t_{3'}=65 \, {\rm meV}$, shown in  
  Fig.~\ref{fig1}b,   reproduces  de Haas -  van Alphen 
 measurements~\cite{Mie97} and
electronic structure calculations~\cite{OMI88}. 
Note, the two pieces of the Fermi surface are
disconnected since $t_3 \neq t_{3'}$.
 
 For the investigation of an electronic pairing state, one has  to 
   determine,   the momentum and frequency dependence of the pairing
 interaction which is responsible for   the magnitude,
$ |\Delta_{l l'}^{i j}|$, of
 the  coordinate space components of the gap function $ \Delta_{l l'}^{i j}  \propto
 \langle c_{i l \uparrow}c_{j l' \downarrow} \rangle  $ and  
the  relative phase relations the $\Delta_{l l'}^{i j} $  establish.
The latter effect   is
  determined by the maximum gain of condensation energy on the
Fermi surface.
 In order to have a   quantitative account for this interplay of magnetic
 correlations, Fermi surface shape and inter- and intra-band coupling,
 I use  
 a  self-consistent summation of bubble and ladder diagrams
 (fluctuation-exchange approximation~\cite{PB94,MS94}) within the
 Nambu-Gorkov description of the
SC state. Both the pairing interaction and the
SC gap function will be determined self consistently.
 Whether     superconductivity
occurs and whether the gap function  possesses nodes on the 
Fermi surface will therefore be the results of our calculation.

The fluctuation-exchange approximation  has been successfully used 
for the investigation of superconductivity in one band  models
relevant for high  temperature  superconductors~\cite{PB94,MS94,GLS96}.
For completeness, I give the set of equations  for
 the two band Hamiltonian, Eq.~\ref{ham},  within  the SC state.
The  effective interactions  of the system,
 \begin{eqnarray}
 V^{\pm} (q)&=& \frac{3 U^2}{2} \left[(1-U \chi^{s}
(q))^{-1} -\frac{1}{3} \right]
  \chi^{  s }( q )  \nonumber \\ & &
\pm \frac{U^2} {2}\left[ (1+U \chi^{  c}( q ))^{-1} -1\right]
  \chi^{   c}( q )
\, ,
\label{Veffsupra}
 \end{eqnarray}
are $(2\times 2)$ matrices in the dimer representation with
particle-hole bubble
   $\chi_{ll'}^{s (c)} (q)= -\sum_k G_{ll'}(k)  G_{l'l}(k+q)  +(-) F_{ll'}(k)  F_{l'l}(k+q) $
for spin- and charge-type excitations. Here $G_{ll'}(k)$  
and $F_{ll'}(k)$  are the normal and anomalous Green's 
functions with $q=({\bf q},i \nu_n)$,
 $k=({\bf k},i \omega_n)$ as well as $\sum_k=T/N \sum_{{\bf k},n}$.  
   ${\bf k}$, ${\bf q}$ are  the two-dimensional wave vectors and
the Matsubara frequencies are   given by
$\omega_n=(2n+1)\pi T$ for fermions and $\nu_n=2n\pi T$ for bosons.
$T$ is the temperature and $N$ the number of unit cells.
From the effective interaction, Eq.~\ref{Veffsupra},  in the 
dimer representation,
one can obtain the intra and inter-band interactions 
$V_{\lambda, \lambda'}^{\pm}(q)$ with $\lambda, \lambda' \in \{ +,- \}$.
The self energy in Nambu representation is   assumed to be  diagonal
in the band representation:  
\begin{equation}
{\hat \Sigma}_{ \lambda}( k) =   Y_{\lambda}(k)  
{\hat \tau^0}+ X_{ \lambda}(k) {\hat \tau^3} +
\Phi_{ \lambda}(k){\hat \tau^1} \, ,
\end{equation}
Here, $Y_{\lambda}(k) \equiv i\omega_n(1-Z_{\lambda}(k)) $
  and $X_{\lambda}(k)$ are the  diagonal   and $\Phi_{\lambda}(k)$
the off -diagonal  components of the self energy in Nambu
representation, respectively, 
and can be determined from the Eliashberg equations,
\begin{eqnarray}
Y_\lambda (k) &=&\sum_{k',\lambda'}  V_{\lambda, \lambda'}^{+}(k-k') 
i\omega_{n'}Z_{\lambda'}(k')/D_{\lambda'}(k')
\nonumber \\
X_\lambda(k) &=&\sum_{k',\lambda'} V_{\lambda , \lambda'}^{+}(k-k') 
 \left(\varepsilon_{{\bf k}' ,\lambda'}+X_{\lambda'}(k')\right)/D_{\lambda'}(k')
\nonumber \\
\Phi_\lambda(k)&=&\sum_{k',\lambda' } \left[V_{\lambda, \lambda'}^{-}(k-k')  +U \right]
\Phi_{\lambda'}(k')/D_{\lambda'}(k')\, ,
\end{eqnarray} 
  with denominator $ D_{\lambda}(k)=(i\omega_n Z_{\lambda}(k))^2-
 (\varepsilon_{{\bf  k},\lambda} +X_{\lambda}(k))^2 -\Phi_{\lambda}(k)^2$.
 The off-diagonal self energy, which signals 
superconductivity,  determines     the gap function  
 $\Delta_{\lambda}(k)=\Phi_{\lambda}(k)/Z_{\lambda}(k)$.
After
analytical continuation to the real axis,$i\omega_n \rightarrow \omega +i0^+$,
this set of coupled equations is solved self consistently  
 using the  numerical  framework described  in  Ref.~\onlinecite{details}.

\begin{figure}
\centerline{\epsfig{file=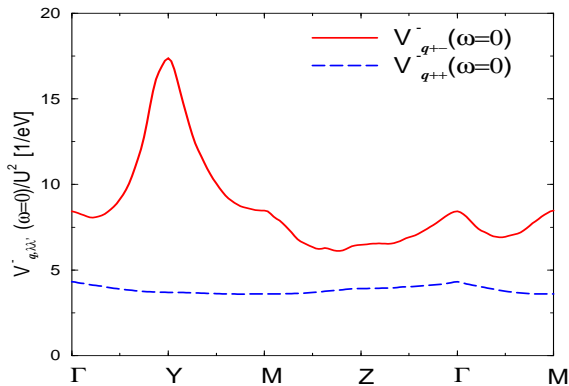,width=8.5cm, height=6cm, scale=0.85}}
\caption{ Effective pairing interaction for intra and inter-band excitations
along the high symmetry lines of the Brillouin zone. Note the pronounced
peak at ${\bf q}=(\pi,0)$ for inter-band excitations.} 
\label{fig2}
\end{figure} 
In Fig.~\ref{fig1}a, we show the unit cell  
 and 
in Fig.~\ref{fig1}b the corresponding Fermi surface, determined by 
$\varepsilon_{{\bf k},\pm}$. The Fermi surface segments of the two bands
appear as one quasi 1-D and one hole pocket part, as seen in  
 magneto-oscillation  experiments~\cite{Caul94,Mie97}.
This Fermi surface   generates $2  {\bf k}_{\rm F}$ intra-band
transitions within the quasi 1-D band. In a real space picture, this corresponds to
a   coupling  between dimers of
the same type. In addition we expect    strong couplings
 $J_{3,3'} \propto t_{3,3'}^2/U$ between dimers of different type.
The numerical analysis of the above set of equations  shows that this latter 
effect, which causes a magnetic  inter-band coupling, dominates.
Correspondingly, the inter-band  effective interactions
 $V^\pm_{{\bf q}+-}(\omega)$
are  large compared to $V^\pm_{{\bf q}++}(\omega)=
V^\pm_{{\bf q}--}(\omega)$
and peaked at ${\bf q}=(\pm \pi,0)$. This result is shown in Fig.2, where we
plot $V^-_{{\bf q}\lambda \lambda'}(\omega=0)$ along  the high symmetry
lines of the Brillouin zone (BZ).
Due to this strong inter-band coupling  we expect a  
relative sign $ -1$   for the gap functions of the two bands:
$\Delta_{{\bf k},+ }\approx  -\Delta_{{\bf k},-}$. Furthermore,  the  pronounced
anisotropy of the pairing interaction, peaked at ${\bf q}=(\pm \pi,0)$,
will cause a substantial momentum
dependence of $\Delta_{{\bf k},\pm}$.

\begin{figure}
\centerline{\epsfig{file=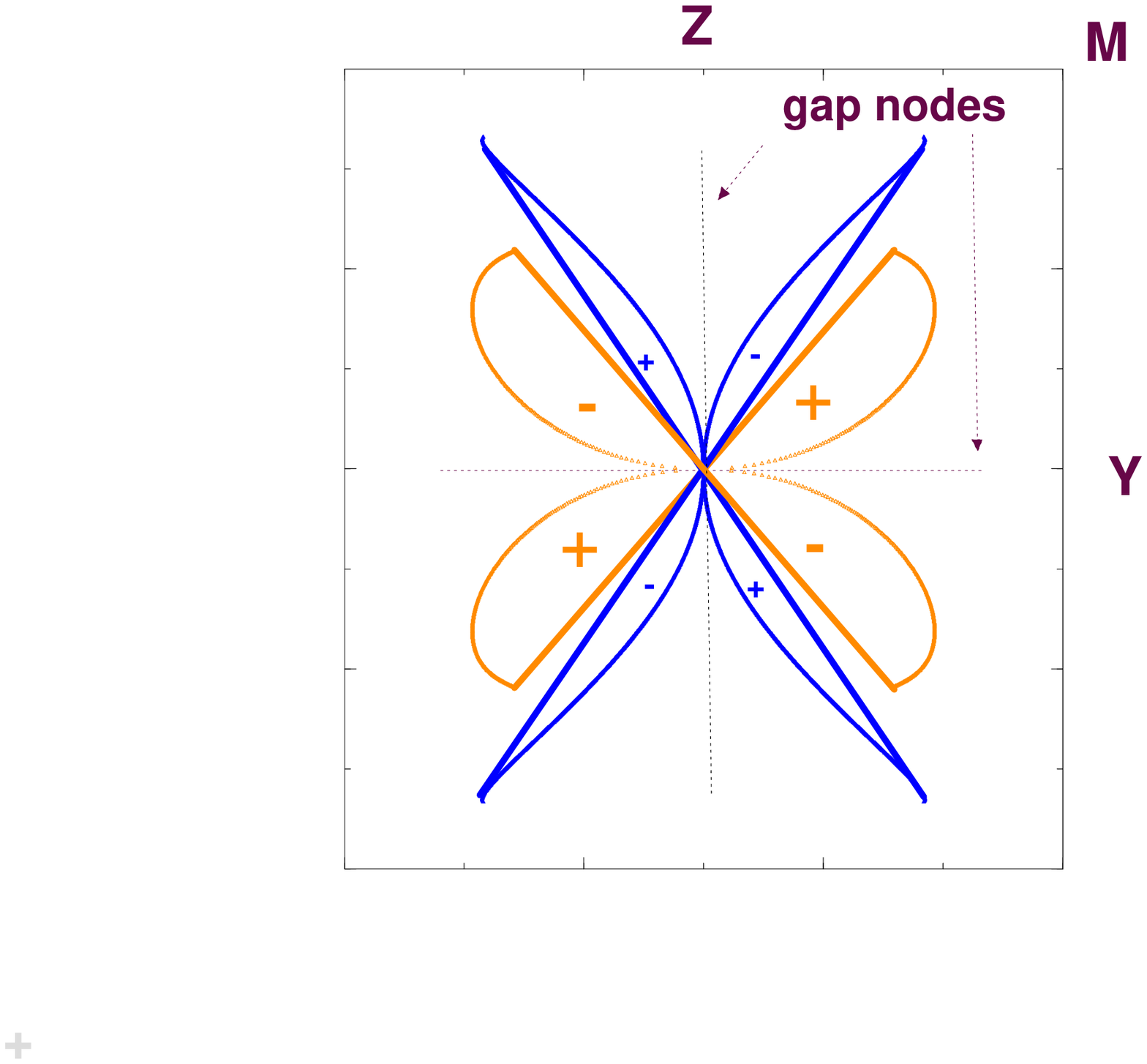,width=8.5cm, height=6cm, scale=0.85}}
\caption{Polar plot of the  superconducting gap functions 
$\Delta_{{\bf k},\pm}$
of the two bands along the Fermi surface. For a given angle
 ${\rm tan}^{-1}(k^{\rm F}_z/k^{\rm F}_y)$ with Fermi vector 
$(k^{\rm F}_y, k^{\rm F}_z)$, the 
amplitude of the gap  corresponds  to the distance from the origin. 
Gap nodes occur   along the $k_y$- and $k_z$-axis.
The suppressed gap close to the diagonal is due to the
absence of a Fermi surface.
 If  $t_3=t_{3'}$ this region vanishes  with equal  gap amplitude 
at the meeting point of the two Fermi surface parts.}
\label{fig3}
\end{figure} 
Our numerical results for the anomalous self energy 
$\Phi_{{\bf k},\pm}(\omega=0) \propto \Delta_{{\bf k},\pm}$ are shown in Fig. 3.
We find a stable SC solution, caused by AF 
spin fluctuations at a temperature $T=8 \, {\rm K}$ which exhibits
zeroth of the gap-function 
  along the lines $k_{y,z}=0$. The leading harmonics
of the gap-function  are given by
$\Delta_{{\bf k},\pm}\approx \Delta^0_{\pm}\sin(k_y/2)\sin(k_z/2)$
with $\Delta^0_+=-0.9 \, {\rm meV}$ and $\Delta^0_-=1.5 \, {\rm meV}$.
 For  $T=15 \, {\rm K}$, within our numerical accuracy,  the
gap function vanishes, consistent with  an estimate for
the  superconducting transition temperature of the order of 
$10 - 15 \, {\rm K}$,
in fair agreement with the experiment.
As shown in Fig. 1c,  this gives 
gap nodes for both pieces of the Fermi surface. 
The symmetry of this superconducting state is d$_{xy}$ which, in addition, 
is out of phase between the two parts of the Fermi surface.

The physical origin of this pairing state and the close resemblance to
the d$_{x^2-y^2}$ state in cuprates can be understood if, for 
illustrating
purposes, one
makes the assumption $t_3=t_{3'}$ for the hopping elements between
different dimer types. 
Now, the two dimer types, which differ  in their   orientation
of the molecule pairs, are indistinguishable and the unit cell reduces  
to that indicated  by dotted lines in Fig.~\ref{fig1}a.
Assuming furthermore that the dimers are arranged on a square lattice,
the BZ is doubled and rotated by $\pi/4$.  
 The resulting Fermi surface is
shown in Fig.~\ref{fig1}c. One easily recognizes that the two branches of the
 Fermi surface of Fig.~\ref{fig1}b correspond, after down-folding into
the reduced BZ,  to states  separated by the dashed line in 
Fig.~\ref{fig1}c.
Since in coordinate space the dominating  AF 
coupling is between nearest neighbors,
it causes peaks in the magnetic susceptibility around $(\pm \pi,\pm \pi)$
of the extended BZ. A quantitative calculation shows that 
the additional anisotropy due to the hopping element $t_2$ causes the
peaks at $\pm (\pi ,\pi)$ to differ from those at $\pm (\pi ,-\pi)$.
Nevertheless, $(\pm \pi,\pm \pi)$ are the dominating momentum transfers
causing  anomalies for {\em hot}  quasiparticles, located on 
Fermi surface  segments   close
the corresponding magnetic zone 
boundary (dashed line in Fig.~\ref{fig1}c)~\cite{HR,SPS98},
which  lead to the    inter-band  coupling in the original BZ of
Fig.~\ref{fig1}b.
On the other hand, as discussed in detail in Refs.~\cite{DP,DJS},
a momentum transfer $(\pm \pi,\pm \pi)$ causes
a superconducting state with gap function $\Delta_{\bf k} \propto
\cos k_y - \cos k_z$.  Rotation by $\pi/4$ and down folding  
into the reduced BZ of Fig.~\ref{fig1}b   yields   
$\Delta_{{\bf k},\pm} \propto \pm   \sin (k_y/2)\sin(k_z/2)
$ which   covers the  general  behavior of the numerical results
presented in Fig.3.
Thus, subject to the modifications brought about by the gap between
 the two branches
of the Fermi surface and the additional anisotropy of the underlying lattice,
the  origin of the pairing state in cuprates and organics  is very 
similar. Due to the important role played by {\em hot} momentum
states for various normal state anomalies of cuprates~\cite{HR,SPS98},
we   expect that they are responsible for normal state anomalies 
in  organics as well.

In conclusion,  I have   shown that spin fluctuations 
are a promising candidate  for the  pairing interaction of 
the superconducting state of organic  molecular crystals.
I find a transition at $10-15 \, {\rm K}$ to a superconducting state, in which 
 the   calculated   occurrence of nodes on the Fermi surface 
is in agreement  with various experimental observations.
Furthermore,  a close similarity of the origin of this pairing symmetry to
the one in cuprate superconductors  was demonstrated. Therefore
it is tempting to conclude that the occurrence of two distinct classes of
quasiparticles, {\em hot} and {\em cold}, depending on the strength of
the effective interaction,   shown to be vitally  important for  the
 normal state anomalies of cuprates,   are    essential
for the corresponding unconventional  behavior    in  
 organics as well.
This would demonstrate that the spin fluctuation model
provides a  general scenario for systems characterized by 
antiferromagnetically short ranged and over-damped spin excitations,
independent of  the specific details  causing these collective spin modes.
Nevertheless, it should be noted that 
 the self consistent weak-coupling approach, used in the present paper,
does not cover  strong coupling effects like the  formation of a 
frustration-induced spin liquid state nor does it provides the 
 desirable   quantitative description of the low-frequency spin
dynamics as reached in cuprates~\cite{DP}; rather it is only  expected to 
give qualitatively correct information about the occurrence and symmetry of
a SC state. 

This work has been supported in part by the Science and Technology
Center for Superconductivity through NSF-grant DMR91-20000,
by the Center for Nonlinear Studies at Los Alamos National Laboratory,
and by the Deutsche Forschungsgemeinschaft.
I thank  R. Gianetta,   C. P. Slichter,    D. Smith, 
D. J.  Van Harlingen and B. Yanoff  and  in particular D. Pines
for helpful discussions.


\end{multicols}

\begin{thebibliography}{99}
\bibitem{alpha} The possibility of superconductivity due to  a spin fluctuation
mechanism in the $\alpha$ and $\beta$ configuration of (ET)$_2$X
will be discussed elsewhere.
\end{thebibliography}
\end{document}